\documentclass[aps,epsfig,prb,preprint]{revtex4}

\usepackage{amsmath}
\usepackage{graphicx}

\def\nn{\nonumber}

\begin{document}

\title{Topological Raman Band in Carbon Nanohorn}

\author{Ken-ichi Sasaki}
\email{sasaki.kenichi@lab.ntt.co.jp}
\affiliation{NTT Basic Research Laboratories, 
Nippon Telegraph and Telephone Corporation,
3-1 Morinosato Wakamiya, Atsugi, Kanagawa 243-0198, Japan}

\author{Yoshiaki Sekine}
\affiliation{NTT Basic Research Laboratories, 
Nippon Telegraph and Telephone Corporation,
3-1 Morinosato Wakamiya, Atsugi, Kanagawa 243-0198, Japan}

\author{Kouta Tateno}
\affiliation{NTT Basic Research Laboratories, 
Nippon Telegraph and Telephone Corporation,
3-1 Morinosato Wakamiya, Atsugi, Kanagawa 243-0198, Japan}


\author{Hideki Gotoh}
\affiliation{NTT Basic Research Laboratories, 
Nippon Telegraph and Telephone Corporation,
3-1 Morinosato Wakamiya, Atsugi, Kanagawa 243-0198, Japan}

\date{\today}

\begin{abstract}
 Raman spectroscopy has been used in chemistry and physics 
 to investigate the fundamental process 
 involving light and phonons (quantum of lattice vibration).
 The carbon nanohorn introduces a new subject to Raman
 spectroscopy, namely topology.
 We show theoretically that a photo-excited carrier with a non-zero
 winding number activates a topological $D$ Raman band through
 the Aharonov-Bohm effect.
 The topology-induced $D$ Raman band can be distinguished from the
 ordinary $D$ Raman band for a graphene edge by its peak position.
\end{abstract}

\maketitle

\begin{figure}[htbp]
 \begin{center}
  \includegraphics[scale=0.7]{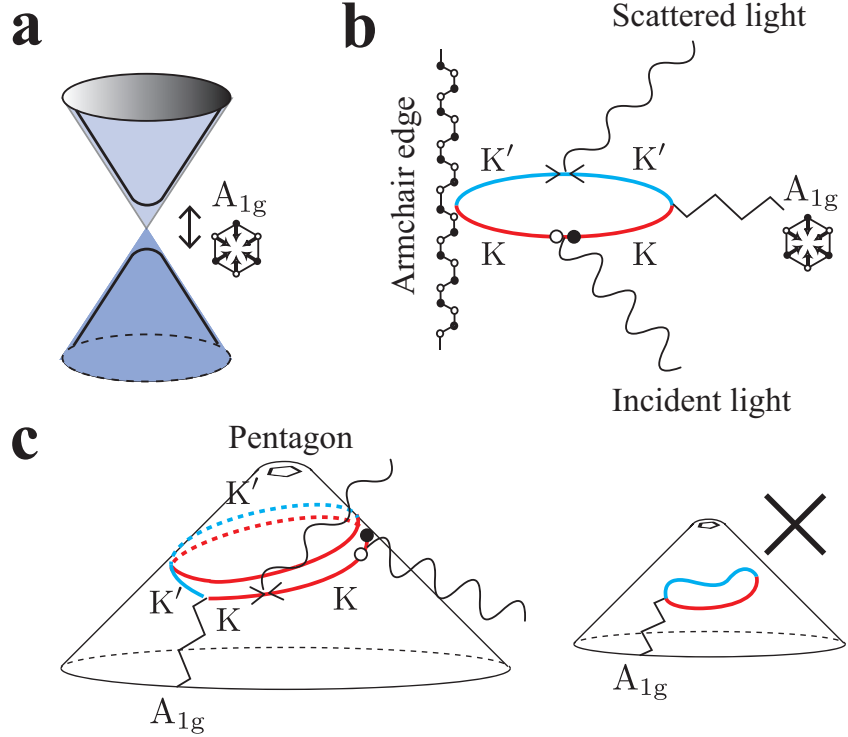}
 \end{center}
 \caption{
 (a) The zone-boundary $A_{1g}$ mode is a gap-opening mode.
 (b) A process exciting the normal $D$ band at an armchair edge. 
 Two changes in valleys are necessary for the activation of the $D$ band
 in the Raman spectrum. (c) A process exciting a topological $D$ band in a
 nanohorn. (inset) A trajectory that does not contribute to the $D$ band
 in a nanohorn.
 }
 \label{fig:point}
\end{figure}

Five-membered rings or pentagons are found throughout the honeycomb network of carbon.
For example, pentagons appear in a fullerene
(buckyball),
at the apexes of carbon nanohorns, at the junctions of carbon nanotubes, and in a flat
sheet of graphene as a constituent of the Stone-Wales
defect.~\cite{kroto85_c60,Iijima92_pentag,iijima99,hashimoto04}
A pentagon is a topological defect, which is represented as the flux of
a pseudomagnetic field pointing perpendicular to the graphene
layer.~\cite{gonzalez92,lammert00,sasaki05,jackiw07}
An interesting consequence of such a flux in quantum mechanics
is the Aharonov-Bohm (AB) effect.
However, the AB effect is usually observed under very silent conditions 
to maintain coherence, which prevents us from utilizing the
AB effect in practical applications.
In this letter, we show that
a topological defect causes a special band (peak) in the Raman
spectrum of a carbon nanohorn,
which we call a topological Raman band.
A topological Raman band is excited through the AB effect, and can be
observed without the need for silent conditions.
A photo-excited ``relativistic'' carrier with a non-zero winding number is the key 
to activating a topological $D$ Raman band.
The topological $D$ band consists of zone-boundary $A_{1g}$ lattice
vibration modes, as well as the normal $D$ band excited near the edge.~\cite{tuinstra70} 
The phonon modes tend to open an energy gap in the Dirac cone by lifting
the degeneracy at the Dirac point (Fig.~\ref{fig:point}(a)).  
We will show that a topological $D$ band is the result of a hybrid
between a pentagon and a Dirac point.
Note that the Dirac point is a topological defect in the Brillouin zone, 
and the topological aspect has been highlighted in the absence of
a backward scattering mechanism leading to the high mobility of metallic
carbon nanotubes.~\cite{ando98}

First, by referring to the Raman process near the armchair edge 
in Fig.~\ref{fig:point}(b), 
we explain that two intervalley scatterings are necessary for the
activation of a $D$ band.
The process starts from an electron-hole pair created by an incident laser light.
Suppose the electron ($\bullet$) and hole ($\circ$) are
located near the K point in the Brillouin zone. 
When the photo-excited electron emits an $A_{1g}$ mode,
the valley changes from K to K$'$ due to momentum conservation.
Meanwhile, the photo-excited hole changes its valley from K to K$'$ as a
result of the intervalley scattering at the armchair edge.~\cite{canifmmode04sec}
After the two scattering events, the hole and electron can be annihilated by
a scattered light emission.
The role of the edge is more clearly understood when we assign different
colors to different valleys; red (blue) is used for K (K$'$).
Namely, a change in colors caused by phonon emission must be
compensated by intervalley scattering at an armchair edge.
In contrast, the color of the trajectory is not altered 
by pair creation or annihilation because the wavelength of a light is
much longer than the wavelength of an electron and hole, 
and optical transitions are possible only when
the valleys of the electron and hole are the same.

Momentum conservation in a nanohorn allows
the activation of a $D$ band even in the absence of an armchair edge.
Figure~\ref{fig:point}(c) illustrates a typical process 
that causes a topological $D$ band in a nanohorn.
The characteristic of the trajectory is
that it revolves twice around the apex.
In contrast, a trajectory that does not turn around the apex 
(inset in Fig.~\ref{fig:point}(c)) does not cause a $D$ band because
such a process does not satisfy momentum conservation as regards emitting an $A_{1g}$ phonon.
The difference between the processes in Fig.~\ref{fig:point}(b) and (c)
is the winding number, which is an integer representing the total number
of times that a curve travels clockwise around the pentagon.
An important point here is that after that the red curve rotates
once around the apex, 
the color changes to blue (i.e., the valley changes automatically)
for a topological reason that will be explained below.

\begin{figure}[htbp]
 \begin{center}
  \includegraphics[scale=1.0]{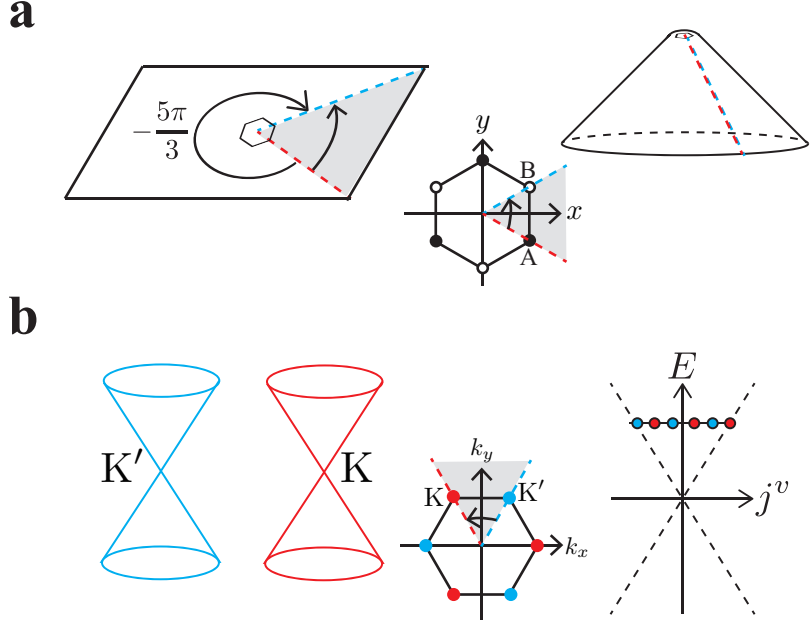}
 \end{center}
 \caption{(a) Sublattice mixing in a nanohorn. 
 A and B atoms are attached so that there is no globally consistent
 definition of sublattices.
 (b) Valley mixing in a nanohorn.
 }
 \label{fig:bzfolding}
\end{figure}

A nanohorn can be obtained by first removing 
the shaded part enclosed by the red and blue
dashed lines shown in Fig.~\ref{fig:bzfolding}(a) from a flat graphene
sheet, and then attaching the red line to the blue line
so that the A-atoms (B-atoms) on the red line are identified with the
B-atoms (A-atoms) on the blue line.
This identification means that it is impossible to make a global
distinction between the A and B sublattices in a nanohorn. 
Furthermore,
due to the removal of the part from the graphene layer,
the corresponding part is also removed from the Brillouin zone,
and the K and K$'$ points are mixed in a nanohorn as shown in Fig.~\ref{fig:bzfolding}(b).
Thus, in a nanohorn, it is also impossible to define the K and K$'$
valleys globally.
These properties, the lack of a global distinction between A and B atoms,
and between K and K$'$ valleys, are unique to a nanohorn and can be
traced back to the existence of a pentagon at the apex.

These features of a nanohorn are simply represented by 
the boundary condition of the wave function:
\begin{align}
 \begin{pmatrix} 
  \psi^{\rm K}_{\rm A}({\bf r}') \cr 
  \psi^{\rm K}_{\rm B}({\bf r}') \cr
  \psi^{\rm K'}_{\rm A}({\bf r}') \cr 
  \psi^{\rm K'}_{\rm B}({\bf r}') 
 \end{pmatrix}
 =  
 \begin{pmatrix} 
  0 & 0 & 0 & -\omega e^{-i\varphi} \cr 
  0 & 0 & -\overline{\omega}e^{-i\varphi} & 0 \cr 
  0 & -\overline{\omega}e^{i\varphi} & 0 & 0 \cr
  -\omega e^{i\varphi} & 0 & 0 & 0 
 \end{pmatrix}
 \begin{pmatrix} 
  \psi^{\rm K}_{\rm A}({\bf r}) \cr 
  \psi^{\rm K}_{\rm B}({\bf r}) \cr
  \psi^{\rm K'}_{\rm A}({\bf r}) \cr 
  \psi^{\rm K'}_{\rm B}({\bf r}) 
 \end{pmatrix},
 \label{eq:bc1}
\end{align}
where ${\bf r}$ is a vector on the surface of a nanohorn, 
$\omega \equiv e^{i\frac{\pi}{3}}$, and the phase $\varphi$ depends on the
position of a pentagon ${\bf R}$ as $\varphi=2{\bf k}_{\rm F}\cdot {\bf R}$
where ${\bf k}_{\rm F}=(\frac{4\pi}{3a},0)$ is the Fermi wave vector at the K
point and $a$ is a lattice constant.~\cite{matsumura98}
The vector ${\bf r'}$ is the position that is given by the
rotation of ${\bf r}$ around the apex: ${\bf r'} =
R(-\frac{5\pi}{3}) {\bf r}$, where $R$ denotes a rotation operator
around the pentagon.
When an electron rotates around ${\bf R}$, 
the amplitude of an A-atom at the K valley, $\psi^{\rm K}_{\rm A}({\bf r})$,
changes to that of a B-atom at the K$'$ valley, $\psi^{\rm K'}_{\rm
B}({\bf r}')$,
with a phase factor $-\omega e^{i\varphi}$.
The valley and sublattice indexes change topologically through a rotation.

The basis of a wave function that can diagonalize the boundary condition
is useful for studying a nanohorn. 
We apply the following unitary transformation to Eq.~(\ref{eq:bc1}), 
\begin{align}
 \begin{pmatrix} 
  \psi_1({\bf r}) \cr 
  \psi_2({\bf r}) \cr
  \psi_3({\bf r}) \cr 
  \psi_4({\bf r})
 \end{pmatrix}
 =
 U^\dagger 
 \begin{pmatrix} 
  \psi^{\rm K}_{\rm A}({\bf r}) \cr 
  \psi^{\rm K}_{\rm B}({\bf r}) \cr
  \psi^{\rm K'}_{\rm A}({\bf r}) \cr 
  \psi^{\rm K'}_{\rm B}({\bf r})
 \end{pmatrix}
\equiv
 \begin{pmatrix}
   e^{i\varphi} & 0 & 0 & 1 \\
 -e^{i\varphi} & 0 & 0 & 1 \\
 0 & e^{i\varphi} & 1 & 0 \\
 0 & -e^{i\varphi} & 1 & 0
 \end{pmatrix}
  \begin{pmatrix} 
  \psi^{\rm K}_{\rm A}({\bf r}) \cr 
  \psi^{\rm K}_{\rm B}({\bf r}) \cr
  \psi^{\rm K'}_{\rm A}({\bf r}) \cr 
  \psi^{\rm K'}_{\rm B}({\bf r})
 \end{pmatrix},
\end{align}
and have 
\begin{align}
 \begin{pmatrix} 
  \psi_1({\bf r}') \cr 
  \psi_2({\bf r}') \cr
  \psi_3({\bf r}') \cr 
  \psi_4({\bf r}') 
 \end{pmatrix}
 =  
 \begin{pmatrix}
  -\omega & 0 & 0 & 0 \cr
  0 & \omega & 0 & 0 \cr
  0 & 0 & -\overline{\omega} & 0 \cr
  0 & 0 & 0 & \overline{\omega}
 \end{pmatrix}
 \begin{pmatrix} 
  \psi_1({\bf r}) \cr 
  \psi_2({\bf r}) \cr
  \psi_3({\bf r}) \cr 
  \psi_4({\bf r}) 
 \end{pmatrix}.
 \label{eq:bc2}
\end{align}
The Hamiltonian for the original wave function is given by the
massless Dirac equation; 
\begin{align}
 H_0 = v_{\rm F}
 \begin{pmatrix}
  0 & \hat{p}_x - i\hat{p}_y & 0 & 0 \cr
  \hat{p}_x + i\hat{p}_y & 0 & 0 & 0 \cr
  0 & 0 & 0 & -\hat{p}_x - i\hat{p}_y \cr
  0 & 0 & -\hat{p}_x + i\hat{p}_y & 0
 \end{pmatrix},
\end{align}
where $\hat{p}_i=-i\hbar \partial_i $ denotes a momentum operator
and $v_{\rm F}$ is the Fermi velocity.
The Hamiltonian for the new wave function becomes
\begin{align}
 U^\dagger
 H_0 U = v_{\rm F}
 \begin{pmatrix}
  0 & 0 & 0 & -\hat{p}_x + i\hat{p}_y \cr
  0 & 0 & -\hat{p}_x + i\hat{p}_y & 0 \cr
  0 & -\hat{p}_x - i\hat{p}_y & 0 & 0 \cr
  -\hat{p}_x - i\hat{p}_y & 0 & 0 & 0
 \end{pmatrix}.
 \label{eq:h2}
\end{align}
Here we define the two-component wave functions $\psi^{v=\pm}$ as
\begin{align}
 \psi^{+} \equiv 
 \begin{pmatrix}
  \psi_2 \cr \psi_3
 \end{pmatrix}, \ \ 
 \psi^{-} \equiv 
 \begin{pmatrix}
  \psi_1 \cr \psi_4
 \end{pmatrix}.
\end{align}
Using Pauli matrices for sublattices, $\psi^v$ is expressed as
$\psi^v = -v e^{i\varphi} \sigma_z \psi^{\rm K}+\sigma_x \psi^{\rm K'}$.
This expression shows that $v=\pm1$ originates from the valley degrees of
freedom.
Since the interaction with the vector potential $A_i$ of light 
is given by replacing $\hat{p}_i$ with $\hat{p}_i-eA_i$ in Eq.~(\ref{eq:h2}),
a laser light cannot induce a $v$-changing transition.

From Eqs.~(\ref{eq:bc2}) and (\ref{eq:h2}),
the wave functions satisfy the same energy eigen equation
\begin{align}
 & E \psi^v({\bf r}) = i\hbar v_{\rm F}
 \begin{pmatrix}
  0 & \partial_x - i\partial_y \cr
  \partial_x + i\partial_y & 0
 \end{pmatrix}
 \psi^v({\bf r}),
\end{align}
with different boundary conditions,
$\psi^v({\bf r}') =v\sigma_z e^{i\frac{\pi}{3}\sigma_z}\psi^v({\bf r})$.
The solution of the eigenvalue equation was constructed by Lammert and
Crespi in a polar coordinate system ${\bf r}\equiv (r,\theta)$ as~\cite{lammert00}
\begin{align}
 \psi^v_{s,k,j^v}(r,\theta) =  N
 e^{i j^v \theta}
 \begin{pmatrix}
  e^{-i \frac{\theta}{2}} J_{|j^v-\frac{1}{2}|}(kr) \cr
  -i s e^{i\frac{\theta}{2}} J_{|j^v+\frac{1}{2}|}(kr)
 \end{pmatrix},
 \label{eq:sol}
\end{align}
where $J_\nu$ is a Bessel function with order $\nu$ and $N$ is a
normalization constant.
The solution is characterized by the band index $s=\pm 1$, the magnitude
of the wave vector $k$, and angular momentum $j^v$, where
the $j^{v}$ values are quantized as~\cite{lammert00}
\begin{align}
 j^{v}=(6/5)(n+v/4),
\end{align}
where $n$ is an integer.
The energy eigenvalue is $sE_k$ (where $E_k = \hbar v_{\rm F}k$), 
and degeneracy is represented by different $j^v$ values.
Because the normalization of $r$ ($0 \leq r \leq R$)
imposes a constraint $-kR \alt j^v \alt kR$,
the density of states increases linearly with increasing energy (See Fig.~\ref{fig:bzfolding}(b)).
Although the index $v$ originates from the valley degrees of freedom,
the actual dependence of $v$ on the wave function appears through the
angular momentum $j^v$.
Namely, the degree of freedom for two valleys is now taken into account
by a shift in the angular momentum: $(5/6)(j^+-j^-) = 1/2 \pmod{1}$.
Hereafter we write $\psi_{s,k,j^v}(r,\theta)$ by omitting the superscript from $\psi^v_{s,k,j^v}(r,\theta)$.

An $A_{1g}$ phonon can be excited without changing $v$.
This is understood by the unitary transformation of the
electron-phonon interaction that is derived for a flat graphene
sheet as 
\begin{align}
 & U^\dagger  
 \begin{pmatrix}
  0 & 0 & 0 & m e^{i {\bf q}\cdot {\bf r}}  \cr
  0 & 0 & m e^{i {\bf q}\cdot {\bf r}}  & 0 \cr
  0 & m e^{-i{\bf q}\cdot {\bf r}}  & 0 & 0 \cr
  m e^{-i{\bf q}\cdot {\bf r}} & 0 & 0 & 0
 \end{pmatrix}
 U \nn \\
 &= 
 \begin{pmatrix}
  m\cos\left(\varphi+{\bf q}\cdot {\bf r}\right) & im \sin\left(\varphi+{\bf q}\cdot {\bf r}\right) & 0 & 0 \cr
  -im\sin\left(\varphi+{\bf q}\cdot {\bf r}\right) & -m \cos\left(\varphi+{\bf
  q}\cdot {\bf r}\right) & 0 & 0 \cr
  0 & 0 & m\cos\left(\varphi+{\bf q}\cdot {\bf r}\right) & im\sin\left(\varphi+{\bf q}\cdot {\bf r}\right) \cr
  0 & 0 & -im\sin\left(\varphi+{\bf q}\cdot {\bf r}\right) & -m\cos\left(\varphi+{\bf q}\cdot {\bf r}\right)
 \end{pmatrix},
 \label{eq:uD}
\end{align}
where $m$ is an electron-phonon coupling and ${\bf q}$ is the wave
vector of an $A_{1g}$ mode measured from the K
point.~\cite{sasaki08ptps,sasaki12_raman}
The $v$-preserving electron-phonon interactions appear at the diagonal
components, which are rewritten as
\begin{align}
 V_{\bf q}^v  = -v m 
 \sigma_z \cos \left(\varphi + {\bf q}\cdot {\bf r} \right),
\end{align}
where $\sigma_z$ results from the fact that an $A_{1g}$ mode opens an
energy gap (Fig.~\ref{fig:point}(a)).
Some $v$-changing interactions appear in the off-diagonal components on
the right-hand side of Eq.~(\ref{eq:uD}), and these do not contribute to
a Raman process. 
Since a $v$-changing optical transition cannot be induced by a laser light,
a $v$-changing scattering by an $A_{1g}$ mode does not satisfy the momentum
selection rule of a Raman process.

The $\sigma_z$ matrix of $V_{\bf q}^v$ plays a decisive role in
determining the ${\bf q}$ value.
First, let us study the following electron-phonon matrix element,
\begin{align}
 M^{(0)}_{elph}
 &\equiv \iint \psi_{s,k,j^v}^\dagger (r,\theta)
 V_{\bf q}^v \psi_{s,k,j^v} (r,\theta) dS \nn \\
 &= -m \iint \psi_{s,k,j^v}^\dagger (r,\theta)
 \psi_{-s,k,j^v} (r,\theta) 
 \cos(\varphi + qr)dS 
 \nn \\
 & \simeq
 \begin{cases}
 -\frac{m}{2} \cos(|j^v| \pi + \varphi) & q= 2k \\
  0 & {\rm otherwise}.
 \end{cases}
\end{align}
Figure~\ref{fig:point}(c) is the process 
described by $M^{(0)}_{elph}$, which does not include 
the effect of the existence of a pentagon.
Since $V_{\bf q}^v$ contains $\sigma_z$, an intraband electronic
scattering induced by the emission of an $A_{1g}$ mode is regarded as an
interband scattering by a potential $\cos(\varphi + qr)$, as shown by the second line. 
This suggests that $M^{(0)}_{elph}$ is suppressed in general.
The last line was obtained by using the asymptotic forms for
non-negative $\alpha$: $J_\alpha(kr)=\sqrt{2/\pi kr}\cos(kr
-\alpha\pi/2-\pi/4)+{\cal O}((kr)^{-1})$.
Although $M^{(0)}_{elph}$ can be of the order of $m$ when $q=2k$, 
$M^{(0)}_{elph}$ does not contribute to the $D$ band because
the matrix element depends on $j^v$, and so each process 
experiences a destructive interference in the Raman process, that is, 
$\sum_{j^v} \cos(|j^v| \pi + \varphi) \approx 0$.
Next, we calculate the following matrix element:
\begin{align}
 M^{(1)}_{elph} 
 &\equiv \iint 
 \psi_{s,k,j^v}^\dagger (r,\theta)
 V_{\bf q}^v
 \psi_{s,k,j^v}(r,\theta-\frac{5\pi}{3}) dS
 \nn \\
 &= \iint 
 \psi_{s,k,j^v}^\dagger (r,\theta)
 V_{\bf q}^v \left[ v \sigma_z e^{i\frac{\pi}{3}\sigma_z}\right]
 \psi_{s,k,j^v} (r,\theta) dS
 \nn \\
 &=-m \cos\left(\frac{\pi}{3}\right)
 \iint
 \rho_{k,j^v}(r)\cos \left(\varphi + qr \right)dS
 + i  
 \sin \left(\frac{\pi}{3}\right) M^{(0)}_{elph},
\end{align}
where $\rho_{k,j^v}(r) \ge 0$ is the probability density.
Figure~\ref{fig:point}(b) is the process described by $M^{(1)}_{elph}$,
which takes account of the effect of a non-zero winding number of a
photo-excited electron (or hole).
When $q=0$, the first term leads to $-\frac{m}{2}\cos(\varphi)$ 
due to normalization $\iint \rho_{k,j^v}(r)dS=1$.
When $q \ne 0$, the first term is suppressed by the integral about $r$.
The first term is independent of $j^v$, which is in contrast to
$M^{(0)}_{elph}$.
For a general case, we define
$M^{(w)}_{elph} 
 \equiv \iint 
 \psi_{s,k,j^v}^\dagger (r,\theta)
 V_{\bf q}^v
 \psi_{s,k,j^v}(r,\theta-\frac{5\pi}{3}w) dS$,
and obtain
\begin{align}
 \frac{\sum_{j^v}M^{(w)}_{elph}}{\sum_{j^v}}
 =
 \begin{cases}
  -iv m \sin \left(\frac{\pi}{3}w \right) \cos (\varphi) & 
  w= {\rm even}, \\
  -m \cos \left(\frac{\pi}{3}w \right) \cos (\varphi) & w= {\rm odd},
 \end{cases}
\end{align}
where we assume $q= 0$.
Similarly, the optical matrix element is given by
\begin{align}
 M_{opt}^{(w)}(s)
 =
 \begin{cases}
  -ev_{\rm F}A_r \frac{is\pi}{2R}\cos\left(\frac{\pi}{3}w\right)  & w= {\rm even}, \\
  ev_{\rm F}A_r \frac{vs\pi}{2R}\sin\left(\frac{\pi}{3}w\right)  & w= {\rm odd}.
 \end{cases}
\end{align}
Here $A_r$ is the vector potential for a circularly polarized light.

The probability amplitude for a Raman process that can contribute to a
topological $D$ band in a nanohorn is written as
\begin{align}
 M^{(l,m,n)}_\varepsilon= \int dS^3 {\rm Tr} \left[
 G^{(l)}_{\varepsilon-E_L}(r,\theta;r'',\theta'')
 \hat{H}_{sc}
 G^{(m)}_{\varepsilon-\hbar \omega}(r'',\theta'';r',\theta') 
 V_{\bf q}
 G^{(n)}_\varepsilon(r',\theta';r,\theta) 
 \hat{H}_{in}
 \right],
\end{align}
where $E_L$ and $\hbar \omega$ are laser energy and phonon energy, respectively.
Here, $G^{(w)}_\varepsilon(r',\theta';r,\theta) $ is the probability
amplitude that an electron at $(r,\theta)$ with energy 
$\varepsilon$ propagates in a nanohorn and arrives at $(r',\theta')$ after
rotating $w$ times around the apex in a clockwise direction.
Explicitly, it is written as
\begin{align}
 G^{(w)}_\varepsilon(r',\theta';r,\theta)=\sum_{s,k,j^v}
 \frac{\left[ v \sigma_z e^{i\frac{\pi}{3}\sigma_z}\right]^w \psi_{s,k,j^v}(r',\theta')\psi_{s,k,j^v}^\dagger(r,\theta)
 }{\varepsilon - sE_k + i\gamma},
\end{align}
where $\hbar/\gamma$ is the mean lifetime of an electron.
It is reasonable to assume that a higher winding
number does not contribute to the Raman process because 
an electron (or hole) experiences scattering caused by impurities and
defects, and so $G^{(w)}_\varepsilon$ is suppressed.
Thus, we focus on the most important process contributing to the $D$
band, that is, the process shown in Fig.~\ref{fig:point}(b):
\begin{align}
 M^{(0,0,1)}_\varepsilon = 
 {\rm Tr} \left[
 \sum_{s,k,j^v}
 \frac{
 M_{opt}^{(0)}(-s) M_{elph}^{(1)} M_{opt}^{(0)}(s)
 }{(\varepsilon-E_L+sE_{k}+i\gamma)(\varepsilon-\hbar
 \omega-sE_{k}+i\gamma)(\varepsilon-sE_{k}+i\gamma)} 
 \right].
\end{align}
When $\gamma \ll \varepsilon_R$ $(\equiv E_L/2)$, a resonance process
dominates other off-resonant processes, and we obtain
\begin{align}
 M^{(0,0,1)}_{\varepsilon_R} \propto 
 \frac{\pi}{\hbar \omega-2i\gamma} \left\{
 \frac{\varepsilon_R }{\gamma } \left(1-i\frac{\gamma}{\hbar
 \omega}\right)-i \right\}|M_{opt}^{(0)}|^2 M_{elph}^{(1)}.
\end{align}
It is noteworthy that the Raman process is represented as a first-order
Raman process, 
i.e., a photo-excited electron is scattered only once, which is the same
as the $G$ Raman band in flat graphene.~\cite{sasaki11_Dband,sasaki12_raman}

The topological $D$ band 
is distinguished from the normal $D$ band induced by the edge of a
nanohorn
by the peak position in the Raman spectrum, because
the self-energy of the $A_{1g}$ mode increases linearly with increasing
$q$.~\cite{sasaki12_migration,sasaki12_raman}
The wave vector of an $A_{1g}$ mode with a topological origin is
$q\simeq 0$, while that of the normal $A_{1g}$ 
mode is $q \simeq 2k$ (where $2k$ is proportional to $E_L/\hbar v_{\rm F}$).
Thus, a topological $D$ peak appears on the low-energy side of the
peak position of the normal $D$ peak, and 
we estimate the shift to be approximately 50 cm$^{-1}$ when $E_L=1.6$ eV
(wavelength is 750 nm).~\cite{sasaki12_migration,sasaki12_raman}
The shift increases almost linearly with increasing the $E_L$ value (see
Fig.3(c) in Ref.~\onlinecite{sasaki12_migration} for details) because of 
the non-dispersive (dispersive) behavior of the topological (normal) $D$ peak.
The abnormal $q$ value (null $q$) means that 
an $A_{1g}$ phonon is created by the forward scattering of a photo-excited electron.
This is in contrast to the fact that 
an $A_{1g}$ phonon with $q \simeq 2k$ is a consequence of the backward scattering
of a photo-excited electron near the armchair edge.~\cite{sasaki12_raman} 
The forward scattering results from 
the identification of A and B atoms, which is enforced through a rotation
about the pentagon and is not seen at the armchair edge.

The region in which a topological $D$ band can be activated, 
is limited to near the pentagon.
This is because, 
the period of a rotation for an electronic state that is distant from
the pentagon is longer than that for a state near the
pentagon, and a state with a long rotational period 
is subject to a strong dephasing effect.
Note that the analysis using the wave function of Eq.~(\ref{eq:sol}) 
is valid when the mean lifetime of electron $\tau$ 
is longer than the period of a rotation,
$\tau_c \equiv 5\pi r/3v_{\rm F}$.
For example, when $\tau=200$ fs, $r< 40$ nm.
There are some perturbations that can shorten $\tau$ or $r$, 
in addition to the $v$-changing part in the electron-phonon interactions
of an $A_{1g}$ mode.
For example, it can be shown that 
the electron-phonon interactions for optical and acoustic phonons near the
$\Gamma$ point and the hybridization between $\sigma$ and $\pi$ orbitals
caused by the curvature at the apex of a nanohorn
are categorized by $v$-changing perturbations.
This also means that the $G$ band is suppressed where the topological
$D$ band is enhanced.

\begin{figure}[htbp]
 \begin{center}
  \includegraphics[scale=0.6]{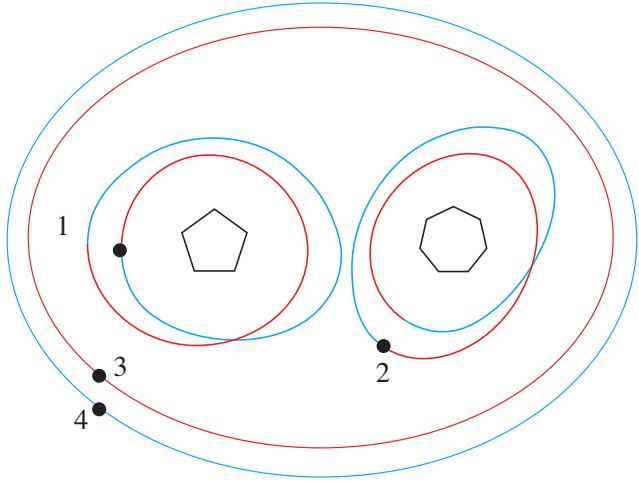}
 \end{center}
 \caption{Paths 1 and 2 are trajectories 
 that are able to activate a topological $D$ band. 
 Paths 3 and 4 that travel around a pentagon-hexagon pair do not
 contribute to the $D$ band. 
 }
 \label{fig:topology}
\end{figure}

A pentagon is not a unique topological defect in a graphene layer.
A heptagon also serves as a topological defect, for which we can derive
the same conclusion as that obtained for a pentagon.
Namely, a topological $D$ band is induced by paths such as 1 and 2
shown in Fig.~\ref{fig:topology}.
There is a pentagon-heptagon pair at the junctions of carbon
nanotubes.~\cite{Iijima92_pentag} 
The trajectories traveling around a pair do not contribute to a
topological $D$ Raman band (paths 3 and 4 in Fig.~\ref{fig:topology}) by cancellation.
In fact, a pentagon (heptagon) is regarded as the flux
$v\Phi_0/4$ ($-v\Phi_0/4$),~\cite{lammert00} 
and the AB effect is suppressed for paths 3 and 4.
We speculate that a topological $D$ band does not vanish 
unless the distance between the pentagon and heptagon is of the order of
the bond length.
This condition would be satisfied for nanohorns and the junctions of
carbon nanotubes, but not for Stone-Wales defects. 

In conclusion, 
a pentagon allows a topological $D$ band to appear in the Raman
spectrum of a nanohorn.
A topological Raman band is the result of the AB effect in Raman spectroscopy,
and a non-zero winding number of the trajectory of a photo-excited electron (or
hole) is the key factor as regards enhancing the intensity. 
The peak position of a topological $D$ band differs by about 50
cm$^{-1}$ from that of the normal $D$ band activated at the edge.
This difference arises due to the lack of a global distinction between A
and B atoms in a nanohorn.

\acknowledgements

We are grateful to Yasuhiro Tokura for helpful discussions.

\bibliographystyle{apsrev}
%
%

\end{document}